\documentclass[12pt]{article}
\usepackage{amsmath, amssymb}
\usepackage{graphicx}
\usepackage{multirow}
\usepackage{bigstrut}
\usepackage{subfigure}

\renewcommand{\Pr}{\mathbb{P}} 
\DeclareMathOperator{\EV}{\mathbb{E}} 
\DeclareMathOperator*{\esssup}{ess\,sup}

\DeclareMathOperator{\LR}{\Lambda}

\DeclareMathOperator{\RIADD}{RIADD}
\DeclareMathOperator{\ARL}{ARL}
\DeclareMathOperator{\SADD}{SADD}
\DeclareMathOperator{\STADD}{STADD}

\newcommand{\T}{T}
\newcommand{\class}{\Delta_{\gamma}}
\newcommand{\norm}[1]{\left\Vert#1\right\Vert}

\graphicspath{{gfx/}}

\begin{document}

\noindent To appear in {\it Communications in Statistics -- Theory and Methods}, {\bf 38}:16, 3225 -- 3239, 2009.
\vskip 5mm

\noindent NUMERICAL COMPARISON OF CUSUM AND SHIRYAEV-ROBERTS PROCEDURES FOR DETECTING CHANGES IN DISTRIBUTIONS
\vskip 3mm

\vskip 5mm
\noindent George V. Moustakides

\noindent Department of Electrical and Computer Engineering

\noindent University of Patras

\noindent 26500 Rio, Greece

\noindent moustaki@upatras.gr

\vskip 5mm
\noindent Aleksey S. Polunchenko and Alexander G. Tartakovsky

\noindent Department of Mathematics

\noindent University of Southern California

\noindent Los Angeles, CA 90089

\noindent \{polunche,tartakov\}@usc.edu

\vskip 3mm
\noindent Key Words: CUSUM test; Fredholm integral equation of the second kind; numerical analysis; quickest change-point detection; sequential analysis; Shiryaev-Roberts test.
\vskip 3mm

\noindent ABSTRACT

\noindent The CUSUM procedure is known to be optimal for detecting a change in distribution under a minimax scenario, whereas the Shiryaev-Roberts procedure is optimal for detecting a change that occurs at a distant time horizon. As a simpler alternative to the conventional Monte Carlo approach, we propose a numerical method for the systematic comparison of the two detection schemes in both settings, i.e., minimax and for detecting changes that occur in the distant future. Our goal is accomplished by deriving a set of exact integral equations for the performance metrics, which are then solved numerically. We present detailed numerical results for the problem of detecting a change in the mean of a Gaussian sequence, which show that the difference between the two procedures is significant only when detecting small changes.
\vskip 4mm

\noindent 1.   INTRODUCTION

\noindent Quickest (sequential) change-point detection deals with detecting changes in distributions that occur at unknown points in time. The goal is to detect the change as soon as possible after its occurrence, while maintaining a prescribed false alarm level. A sequential change-point detection procedure is defined as a stopping time $\T$ (with respect to an observed sequence $\{X_n\}_{n\ge1}$).

In this paper we consider the simplest version of the change-point detection problem where we assume that the observations are independent and identically distributed (i.i.d.) before the change with a common density $f$ and i.i.d. with a different density $g$ after the change, both of which are considered known. Our goal is to provide a comparative study of the main competitors -- the Cumulative Sum (CUSUM) procedure introduced by Page (1954) and the Shiryaev-Roberts procedure introduced by Shiryaev (1961) for the Brownian motion case and Roberts (1966) for discrete time.

It is known that both schemes enjoy specific optimality properties under different optimality criteria. More precisely, it follows from Moustakides (1986) that the CUSUM procedure is (min-max) optimal with respect to Lorden's (1971) detection measure
$$
\mathcal{J}_{\rm L}(\T)=\sup_{\nu\ge0}\esssup\EV_\nu [(\T-\nu)^+|X_1,\ldots,X_{\nu}]\eqno(1.1)
$$
in the class $\class=\{\T\colon\EV_\infty[\T]\ge\gamma\}$ of detection procedures for which the average run length (ARL) to false alarm $\EV_\infty[\T]$ is no smaller than a given number $\gamma>1$. Hereafter $\EV_\nu$ denotes the operator of expectation when the point of change is $\nu$ ($\nu=\infty$ means that there is no change) and $y^+$ stands for the positive part of $y$. On the other hand, it follows from Pollak and Tartakovsky (2009) that the Shiryaev-Roberts procedure is optimal with respect to the relative integral average detection delay measure
$$
\RIADD(\T)=\frac{\sum_{\nu=0}^\infty\EV_\nu[(\T-\nu)^+]}{\EV_\infty[\T]},\eqno(1.2)
$$
again within the same class $\class$. This measure is also equivalent to the stationary average detection delay when detecting changes occurring at a distant time horizon (see Section 2 for further details). These latter performance measures and their corresponding properties were motivated by similar results obtained for the Shiryaev-Roberts procedure for the continuous-time Brownian motion model; see Shiryaev (1963) and Feinberg and Shiryaev (2006). Finally, we should mention that the two tests are asymptotically optimal as $\gamma\to\infty$ (i.e., for low false alarm rate) with respect to both performance measures $\mathcal{J}_{\rm L}$ and $\RIADD$ and for a class of observation processes that is much richer than the simple i.i.d.~case (see, e.g., Lai, 1998 and Tartakovsky and Veeravalli, 2004).

It is of major practical interest to compare the two popular tests with respect to the two aforementioned measures, since each performance measure attempts to capture completely different change-point scenarios. The exact analytical characterization of the two performance measures was recently made possible by Moustakides et al. (2009) through a set of integral equations. These equations were in turn solved numerically using very simple techniques, yielding the final performance metrics. Due to the corresponding exact optimality properties, it is expected that CUSUM will outperform the Shiryaev-Roberts procedure with respect to Lorden's performance measure $\mathcal{J}_{\rm L}$, while the Shiryaev-Roberts procedure will be superior with respect to the relative integral average detection delay $\RIADD(\T)$. Our goal is to quantify this difference and asses its importance.

Comparisons of the two tests have been performed in the past. Roberts (1966) considered a change in the mean of a Gaussian sequence and the two tests were compared with respect to their ARL to detection $\EV_0[\T]$ value using Monte Carlo simulations. CUSUM was found to be better and this is not surprising since $\EV_0[\T]$, in both tests, coincides with Lorden's measure. Pollak and Siegmund (1985) performed a comprehensive asymptotic study (as $\gamma\to\infty$, i.e., for low false alarm rate) for the problem of detecting a change in the drift of the Brownian motion and found that CUSUM performs better for changes that occur in the beginning (i.e., $\nu=0$), while the Shiryaev-Roberts procedure outperforms CUSUM with respect to the conditional average detection delay $\EV_\nu[\T-\nu|\T>\nu]$ when $\nu\to\infty$. Srivastava and Wu (1993) also presented an asymptotic analysis (as $\gamma\to\infty$) for Brownian motion but for the stationary average detection delay case. Tartakovsky and Ivanova (1992) obtained accurate asymptotic approximations for the ARL to false alarm and the average detection delay for the processes with i.i.d. increments (in continuous and discrete time) and performed a detailed numerical comparison of the CUSUM and Shiryaev-Roberts procedures for an exponential model. Finally, Dragalin (1994) analyzed the CUSUM procedure for the problem of detecting a change in the mean of the normal distribution in terms of the ARL to false alarm $\EV_\infty[\T]$ and the ARL to detection $\EV_0[\T]$, using a precise numerical technique.

Despite the previously mentioned results, a comprehensive comparison of the two tests for the discrete-time model in a non-asymptotic setting, i.e., for arbitrary false alarm rate, is still missing. In the present paper we give a partial answer to this question by proposing a technique that can perform the desired comparison numerically, being however of sufficient generality to include any i.i.d. observation model.

The paper is organized as follows. In Section 2 we provide a brief overview of results in change-point detection, introduce our notation and describe the CUSUM and Shiryaev-Roberts procedures. In Section 3 we derive integral equations for the performance metrics of interest and provide a simple numerical solution that allows for efficient computation of the operating characteristics. In Section 4 we present the results of our numerical methodology in the example of detecting a change in the mean of a Gaussian sequence.

\vskip 3mm

\noindent 2. CHANGE-POINT DETECTION PROCEDURES

\noindent 2.1 Notation and Problem Formulation

\noindent Let a sequence $\{X_n\}_{n\ge 1}$ of independent random variables be observed sequentially. Initially the sequence is ``in-control'', i.e., all observations are coming from the same probability density $f(x)$. At an unknown time instant $\nu\ge0$ something happens and the sequence runs ``out of control'' by abruptly changing its statistical properties so that from $\nu+1$ on the density is $g(x)\not\equiv f(x)$. This change has to be detected as quickly as possible, while controlling false alarms at a given level.

Given the sequence $\{X_n\}_{n\ge 1}$, a sequential detection procedure is identified with a stopping time $\T$ adapted to the filtration $\{\mathcal{F}_n\}_{ n\ge0}$, where $\mathcal{F}_n=\sigma(X_1,\ldots,X_n)$ is the (smallest) $\sigma$-algebra generated by the observations up to time instant $n$, with $\mathcal{F}_0$ denoting the trivial $\sigma$-algebra. In other words, for $n\ge 0$, the set $\{T\le n\}$ belongs to the $\sigma$-algebra $\mathcal{F}_n$. At time instant $\T$ the procedure stops and declares that a change has occurred. The design of quickest change-point detection procedures involves optimizing a tradeoff between two types of performance metrics, one being a measure of the detection delay and the other of  the rate of false alarms. Let us denote with $\Pr_\nu$ and $\EV_\nu$ the probability and the corresponding expectation induced by a change occurring at time $\nu\ge0$. According to this definition $\Pr_\infty$ ($\EV_\infty$) denotes the probability (expectation) when there is no change, while $\Pr_0$ and $\EV_0$ the corresponding quantities when the change takes place before  observations become available.

We are interested in two different mathematical setups. In the first we follow the minimax approach proposed by Lorden (1971) and expressed through (1.1). A similar measure, seemingly less pessimistic (for a discussion see Moustakides, 2008), was proposed in Pollak (1985) where detection speed is expressed via the supremum average (conditional) detection delay
$$
\SADD(\T)=\sup_{0\le\nu<\infty}\EV_\nu[\T-\nu|\T>\nu].\eqno(2.1)
$$
As we have mentioned in the introduction, Lorden (1971) proposed to minimize the measure defined in (1.1) in the class $\class$, i.e., subject to the constraint $\EV_\infty[\T]\ge\gamma$  imposed on the ARL to false alarm. Following the same principle, Pollak (1985) suggested a similar constrained optimization problem with Lorden's measure $\mathcal{J}_{\rm L}(\T)$ replaced by $\SADD(\T)$. We should emphasize that in the case of the two popular tests we have $\mathcal{J}_{\rm L}(T)=\SADD(\T)=\EV_0[\T]$. Consequently, even though we will refer to $\SADD(\T)$ as our first performance measure, one should keep in mind that, at the same time, we refer to Lorden's essential supremum measure as well.

The second formulation aims at minimizing the relative integral average detection delay defined in (1.2) subject to the lower bound on the ARL to false alarm  $\EV_\infty[\T]\ge\gamma$ (i.e., the class $\class$). As has been shown by Pollak and Tartakovsky (2009), this is instrumental in detecting  a change that occurs in the distant future (large $\nu$) and is preceded by a stationary flow of false alarms. Specifically, consider a context in which it is of utmost importance to detect a real change as quickly as possible even at the expense of raising many false alarms (using a repeated application of the same stopping rule) before the change occurs. This essentially means that the change-point $\nu$ is substantially larger than the ARL to false alarm $\gamma$ which, in this case, defines the mean time between (consecutive) false alarms. Let $T_1,T_2,\ldots$ denote sequential independent repetitions of the stopping time $\T$ and let ${\cal T}_j=T_1+T_2+\cdots+T_j$ be the time of the $j$-th alarm. Define $I_\nu=\min\{j\ge1\colon {\cal T}_j>\nu\}$. In other words, ${\cal T}_{\scriptscriptstyle I_\nu}$ is the time of detection of a true change that occurs at $\nu$ after $I_\nu-1$ false alarms have been raised. Write
$$
\STADD(\T)=\lim_{\nu\to\infty}\EV_\nu[{\cal T}_{\scriptscriptstyle I_\nu}-\nu]
$$
for the limiting value of the average detection delay that we will refer to as the {\em stationary average detection delay} (STADD). It follows from Theorem 2 in Pollak and Tartakovsky (2009) that
$$
\STADD(\T)=\frac{\sum_{k=0}^\infty\EV_k[(\T-k)^+]}{\EV_\infty[\T]}=\RIADD(\T).\eqno(2.2)
$$
$\STADD(\T)$ is the second performance measure we will adopt for our comparisons.

We note that the stationary average detection delay measure $\STADD(\T)$ has been first introduced by Shiryaev (1961, 1963) for the problem of detecting a change in the drift of a Brownian motion, where also the Shiryaev-Roberts procedure has been introduced for the first time and shown to be optimal with respect to $\STADD(\T)$ in the class of procedures with $\EV_\infty[\T]=\gamma$. See also Feinberg and Shiryaev (2006).

\vskip0.3cm
\noindent 2.2 CUSUM and Shiryaev-Roberts Procedures

\noindent For $n\ge 1$, define
$$
\LR_n=\frac{g(X_n)}{f(X_n)},
$$
the ``instantaneous" likelihood ratio between the post-change and pre-change hypotheses. To avoid complications we shall assume that $\LR_1$ is continuous. Yet, if need be, the case where $\LR_1$ is non-arithmetic can also be covered with a certain additional effort.

Using the previous notation, the Shiryaev-Roberts procedure stops and raises an alarm at
$$
\T_{A}^{\scriptscriptstyle\mathrm{SR}}=\inf\{n\ge1\colon R_n\ge A\},\quad \inf\{\emptyset\}=\infty,
$$
where $R_n$ is the Shiryaev-Roberts detection statistic defined as
$$
R_n=\sum_{k=1}^{n}\prod_{j=k}^n\LR_j,\eqno(2.3)
$$
and $A=A_\gamma>0$ is a threshold chosen so that the false alarm constraint $\EV_\infty[\T_{A}^{\scriptscriptstyle\mathrm{SR}}]=\gamma$ is met.

It is straightforward to verify from (2.3) that the Shiryaev-Roberts statistic allows for the following convenient recursive representation
$$
R_n=(1+R_{n-1})\LR_n,\quad R_0=0.
$$
Pollak and Tartakovsky (2009) showed that the Shiryaev-Roberts procedure $\T_{A_\gamma}^{\scriptscriptstyle\mathrm{SR}}$ is {\em exactly} optimal in the sense of minimizing the relative integral average detection delay $\RIADD(\T)$ and hence due to (2.2) the stationary average detection delay $\STADD(\T)$ for every $\gamma>1$.

The CUSUM test is motivated by the maximum likelihood argument and is based on the comparison of the maximum likelihood ratio
$$
V_n=\max_{1 \le k \le n}\prod_{j=k}^{n}\LR_k
$$
with a positive threshold $A$, i.e., the CUSUM stopping time is defined as
$$
\T_{A}^{\scriptscriptstyle\mathrm{CS}}=\inf\{n\ge1\colon V_n\ge A\}, \quad \inf\{\emptyset\}=\infty.\eqno(2.4)
$$
It is easily verified that the statistic $V_n$ can be computed recursively as
$$
V_n=\max\{1,V_{n-1}\}\LR_n,\quad V_0=1.\eqno(2.5)
$$

Note that conventional Page's CUSUM statistic is given by
$$
W_n=\max\{0,W_{n-1}+\log\LR_n\},\quad W_0=0.\eqno(2.6)
$$
Clearly, the trajectories of this statistic coincide with the trajectories of $\log V_n$ {\it on the positive half plane} and, therefore, the CUSUM stopping time defined in (2.4) is equivalent to familiar Page's stopping time
$$
T_A^{\scriptscriptstyle\mathrm{PG}}=\inf\{n\ge1\colon W_n\ge \log A\}
$$
whenever $A>1$. Note also that, while not crucial for most practical purposes, the CUSUM procedure given by (2.4) and (2.5) is more general than the classical Page rule since it allows for thresholds $A\le1$ (the classical test with such thresholds stops in one step).

Threshold $A=A_\gamma$ is chosen in such a way that the ARL to false alarm meets the constraint $\EV_\infty[\T_{A_\gamma}^{\scriptscriptstyle\mathrm{CS}}]=\gamma$ exactly. While we use the same notation $A$ for the thresholds in both the CUSUM and Shiryaev-Roberts procedures, to avoid confusion we stress that the thresholds are in fact fairly different for achieving the same false alarm rate.

In the minimax setting, Lorden (1971) proved that CUSUM is asymptotically (as $\gamma\to\infty$) optimal in the sense of  minimizing the ${\cal J}_{\rm L}(\T)$ over all stopping times $\T$ such that $\EV_\infty[\T]\ge\gamma$. This result was later improved by Moustakides (1986) who showed that CUSUM is {\em exactly} optimal for every $\gamma>1$ (for a different proof, see Ritov, 1990).

\vskip 3mm

\noindent 3. INTEGRAL EQUATIONS FOR PERFORMANCE METRICS AND NUMERICAL APPROXIMATIONS

This section is devoted to our analytical methodology as applied to the Shiryaev-Roberts and CUSUM procedures. We follow the technique developed in Moustakides et al. (2009) for the generalized Shiryaev-Roberts procedure which can be initialized from any point $R_0=r\in[0,A]$ and not necessarily from 0 as in the classical case we adopt here.

We recall the important observation mentioned earlier that for both CUSUM and the Shiryaev-Roberts procedure Lorden's essential supremum measure $\mathcal{J}_{\rm L}(T)$ defined in (1.1) and Pollak's supremum measure $\SADD(\T)$ defined in (2.1) are attained at $\nu=0$, that is,
$$
\mathcal{J}_{\rm L}(\T_{A}^{\scriptscriptstyle\mathrm{CS}})=\SADD(\T_{A}^{\scriptscriptstyle\mathrm{CS}}) =\EV_0[\T_{A}^{\scriptscriptstyle\mathrm{CS}}], \quad
\mathcal{J}_{\rm L}(\T_{A}^{\scriptscriptstyle\mathrm{SR}})=\SADD(\T_{A}^{\scriptscriptstyle\mathrm{SR}}) =\EV_0[\T_{A}^{\scriptscriptstyle\mathrm{SR}}],
$$
where $\EV_0[\T]$ is the average detection delay when the change occurs before surveillance begins (also known as the ARL to detection). Therefore, in order to compare these procedures in the worst-case scenario it is sufficient to compute the ARL to detection. Since the CUSUM procedure is optimal with respect to Lorden's measure $\mathcal{J}_{\rm L}(\T)$ in the class $\class$, it is expected that it will perform better than the Shiryaev-Roberts procedure. On the other hand, since the Shiryaev-Roberts procedure is optimal with respect to the stationary average detection delay $\STADD(\T)$, it is expected that it will perform better than the CUSUM procedure when detecting distant changes.

In order to unify the approach for both tests, consider a sequential scheme whose stopping time is of the form
$$
\T_{A}=\inf\{n\ge1\colon S_n\ge A\},\quad\inf\{\emptyset\}=\infty\eqno(3.1)
$$
with the corresponding Markov detection statistic satisfying
$$
S_n=\xi(S_{n-1})\LR_n\,,\quad n=1,2,\ldots,\eqno(3.2)
$$
where $S_0=s\in[0,A]$ is a given (fixed) starting point, $A$ is a positive threshold and $\xi(s)$ is a sufficiently smooth positive-valued (for all $s\in[0,A]$) function.

It is evident that both the CUSUM and Shiryaev-Roberts statistics are of this form. Indeed, for CUSUM $\xi(S)=\max\{1,S\}$ and for the Shiryaev-Roberts procedure $\xi(S)=1+S$. Next, we derive a set of equations for the performance metrics of the generic detection procedure defined in (3.1) and (3.2), which we can then easily adapt to the CUSUM and Shiryaev-Roberts procedures by selecting the appropriate form of $\xi(S)$.

For fixed $A>0$ and $s\in[0,A]$, define $\phi_i(s)=\EV_i[\T_A]$, where $i=\{\infty,0\}$. It is apparent that $\phi_\infty(s)=\EV_\infty[\T_A]$ is the ARL to false alarm and $\phi_0(s)=\EV_0[\T_A]$ is  the ARL to detection. For $k\ge0$ and $s\in[0,A]$, define $\delta_k(s)=\EV_k[(\T_A-k)^+]$ and let $F_i(x)=\Pr_i(\LR_1\le x)$ denote the cumulative distribution function of the likelihood ratio $\LR_1$ for  $i=\{\infty,0\}$.

Using the Markov property of the statistic $S_n$ and the argument of  Moustakides et al. (2009), we obtain
$$
\phi_i(s)=1+\int_0^A\phi_i(x)\left[\dfrac{\partial}{\partial x}F_i\left(\frac{x}{\xi(s)}\right)\right]dx,\eqno(3.3)
$$
and
$$
\delta_k(s)=\int_0^A\delta_{k-1}(x)\left[\dfrac{\partial}{\partial x} F_\infty\left(\dfrac{x}{\xi(s)}\right)\right]dx,\quad k\ge 1\eqno(3.4)
$$
with the initial condition  $\delta_0(s)=\EV_0[\T_A]=\phi_0(s)$ and the latter function satisfying (3.3). The integral equation (3.3) yields the ARL to false alarm $\EV_\infty [\T_A]$ and the ARL to detection $\EV_0[\T_A]$ while (3.4) recursively computes $\EV_k[(\T_{A}-k)^+]$ as functions of the starting point $s\in[0,A]$.

In order to compute the stationary average detection delay $\STADD(\T_A)$ defined in (2.2), we need to evaluate the integral average detection delay $\psi(s)=\sum_{k=0}^\infty\EV_k[(\T_{A}-k)^+]$. According to our previous definitions we observe that
$$
\psi(s)=\sum_{k=0}^\infty\delta_k(s).\eqno(3.5)
$$
To find a more convenient formula for $\psi(s)$, let us
introduce a linear operator associated with the kernel $\mathcal{K}_\infty(x,y)=
\tfrac{\partial}{\partial x}F_\infty\left(\tfrac{x}{\xi(y)}\right),$  which transforms a given function $\zeta$ into a new function $\eta$ as follows
$$
\eta(y)=(\mathcal{K}\circ\zeta)(y)=\int_0^A\zeta(x)\,\mathcal{K}_\infty(x,y)\,dx.
$$
Notice now that $\delta_k(s)$, defined in (3.4), can be seen as the repetitive application of this linear operator onto the function $\delta_0(s)$.
In terms of this operator, equation (3.4) can be rewritten as
$$
\delta_k(s)=(\mathcal{K}_\infty^k\circ\delta_0)(s)
=\underbrace{\int_0^A\cdots\int_0^A}_{\text{$k$ times}}\delta_0(x_0)\,\underbrace{\mathcal{K}_\infty(x_0,x_1)\,dx_0\ldots\mathcal{K}_\infty(x_{k-1},s)\,dx_{k-1}}_{\text{$k$ times}}
$$
with the convention that $(\mathcal{K}_\infty^0\circ\delta_0)(s)=\delta_0(s)$. Consequently, this operator representation of (3.4) enables one to turn (3.5) into the following Neumann series
$$
\psi(s)=\sum_{k=0}^\infty\delta_k(s)=
\sum_{k=0}^\infty(\mathcal{K}_\infty^k\circ\delta_0)(s),
$$
which by the geometric series convergence theorem leads to the following equation
$$
\psi(s)=\delta_0(s)+\int_0^A\psi(x)\left[\dfrac{\partial}{\partial x}F_\infty\left(\dfrac{x}{\xi(s)}\right)\right]dx.\eqno(3.6)
$$
The geometric series convergence theorem applies since the spectral radius of the operator $\mathcal{K}_\infty(x,y)$ is strictly less than 1. The proof of the latter fact for the Shiryaev-Roberts procedure can be found in Moustakides et al. (2009). For the CUSUM procedure the argument is essentially the same.

Note that functions $\phi_i(s)=\phi^\xi_i(s)$ and $\psi(s)=\psi^\xi(s)$ depend on $\xi$. Taking $\xi(s)=\max(1,s)$ and $\xi(s)=1+s$, integral equations (3.3) and (3.6) allow for the following computation of the stationary average detection delay of the CUSUM and Shiryaev-Roberts procedures
$$
\STADD(T_A)=\psi(0)/\phi_\infty(0),
$$
while we recall that the supremum average detection delay $\SADD(\T_A)=\phi_0(0)$ is computed from equation (3.3) with  $\xi(s)=\max(1,s)$ for CUSUM and $\xi(s)=1+s$ for the Shiryaev-Roberts procedure.

Observe that both equations (3.3) and (3.6) for $i=\{\infty,0\}$ are Fredholm equations of the second kind (see, e.g., Petrovskii, 1957 and Kress, 1989). It is known that, provided $1$ is not an eigenvalue of the kernel $\mathcal{K}_i(x,y)=
\tfrac{\partial}{\partial x}F_i\left(\tfrac{x}{\xi(y)}\right)$, these equations possess unique solutions. It is also worth emphasizing that throughout the paper, kernels $\mathcal{K}_i(x,y)$ are sufficiently smooth, because the likelihood ratio was assumed to be continuous.

In general, it is not feasible to obtain analytical solutions since the corresponding integral equations are difficult to solve. Alternatively, we can attempt to solve these equations numerically. Efficient numerical schemes are
developed in Kantorovich and Krylov (1958), Petrovskii (1957) and Atkinson and Han (2001). The most popular approach consists in applying a quadrature rule to approximate the integral appearing on the right-hand side of (3.3) and (3.6). Specifically, once the choice of a quadrature rule is made, the interval $[0,A]$ is divided into a partition $0=x_0<x_1<\ldots<x_N=A$, and the functions $\phi_i(x)$ are sampled at the breakpoints producing column vectors $\boldsymbol{\phi}_i=[\phi_i(x_0),\phi_i(x_1),\ldots,\phi_i(x_N)]'$. The integral is then evaluated using the quadrature rule by the following simple matrix-vector multiplication
$$
\int_0^A\mathcal{K}_i(x,y)\,\phi_i(y)\,dy=\boldsymbol{K}_i\widetilde{\boldsymbol{\phi}}_i +\boldsymbol{\varepsilon} ,
$$
where $\boldsymbol{\varepsilon}$ is the approximation error, $\boldsymbol{K}_i$ is a matrix that depends on the chosen quadrature rule and the partition $\{x_i\},\{y_i\}$, and $\widetilde{\boldsymbol{\phi}}_i=[\widetilde{\phi}_i(x_0),\widetilde{\phi}_1(x_1),\ldots,\widetilde{\phi}_i(x_N)]'$
with $\widetilde{\phi}_i(x)$ denoting the approximation to $\phi_i(x)$.  A similar argument applies to the equation of $\psi(x)$.

Matrices $\boldsymbol{K}_i$ can be found using numerical integration. To this end, we will use the simplest method sampling the interval $[0,A]$ equidistantly at the points $x_j=y_j=jh,~j=0,\ldots,N$ with $h=A/N$ and defining the $(n,m)$-element of matrices $\boldsymbol{K}_i$ of size $N$-by-$N$ as
$$\label{eq:def-matrix-K}
(\boldsymbol{K}_i)_{n,m}=
F_i\left(\dfrac{x_n}{\xi(x_m)}\right)-F_i\left(\dfrac{x_{n-1}}{\xi(x_m)}\right),
\quad 1\le n,m\le N.\eqno(3.7)
$$
Beyond the node points, the unknown function $\phi_i(x)$ is then evaluated as
$$
\widetilde{\phi}_i(x)=1+\sum_{j=0}^N\mathcal{K}_i(x,y_j)\widetilde{\phi}_i(y_j).
$$

Regardless of the specific form of pre and post-change densities, the dominant eigenvalue $\widetilde{\lambda}_{\max}$ of the matrix $\boldsymbol{K}_\infty$ defined by (3.7) for $i=\infty$ is {\em strictly}  less than 1 (and positive). This follows from the following inequality
$$
\widetilde{\lambda}_{\max}\le\norm{\boldsymbol{K}_\infty}_\infty .
$$

Combining all previous observations yields
$$
\widetilde{\boldsymbol{\phi}}_i=
J+\boldsymbol{K}_i\widetilde{\boldsymbol{\phi}}_i,\,\,i=\{\infty,0\},\eqno(3.8)
$$
$$
\widetilde{\boldsymbol{\psi}}=
\widetilde{\boldsymbol{\phi}}_0+\boldsymbol{K}_\infty\widetilde{\boldsymbol{\psi}},\eqno(3.9)
$$
where $\widetilde{\boldsymbol{\phi}}_i=[\tilde{\phi}_i(0),\tilde{\phi}_i(h),\ldots,
\tilde{\phi}_i(A)]'$ and $\widetilde{\boldsymbol{\psi}}=[\tilde{\psi}(0),\tilde{\psi}(h),\ldots,
\tilde{\psi}(A)]'$ with $\widetilde{\phi}_i(x)$ and $\widetilde{\psi}(x)$ denoting the approximations to $\phi_i(x)$ and $\psi(x)$, respectively, and $J=[1,1,\ldots,1]'$.

Linear matrix equations (3.8) and (3.9) constitute a complete set of approximations to their corresponding exact integral counterparts. These equations can be solved either directly or iteratively. Direct methods are known to be more accurate, but the accuracy comes at a price of considerable memory requirements. Iterative methods, although less memory demanding, are less accurate. It is evident that the accuracy of the proposed numerical method strongly depends on the number of sample points $N$: the larger it is, the finer the partition and the more accurate the numerical approximation. Such a conclusion follows from the analysis performed, e.g., in Kantorovich and Krylov (1958) and Atkinson and Han (2001).

Fredholm equations for the ARL to false alarm $\EV_\infty [\T]$ and the ARL to detection $\EV_0[\T]$, but only for the CUSUM procedure, have been previously considered in the literature (see, e.g., Dragalin, 1994 and references therein). These equations rely on the classical form of CUSUM given in (2.6) and, therefore, differ from the ones presented in (3.3). The unified approach we propose here, in addition to the obvious advantage of being applicable to a whole class of procedures that includes the Shiryaev-Roberts test, CUSUM and EWMA (not treated here) as particular cases, also simplifies the computations for CUSUM. Indeed, note that in the conventional approach usually considered in the literature (in particular by Dragalin, 1994), the CUSUM statistic is considered as reflected from the unit barrier\footnote{Here we refer to the exponentially transformed CUSUM statistic $e^{W_n}$, where $W_n$ is given by the recursion (2.6).}, which generates a nonzero probability mass (atom) at 1. Consequently, point 1 requires special treatment, complicating the corresponding integral equations. This drawback disappears under the alternative form (2.5) we adopt here. As we can see, in our approach point 1 has zero probability like any other point in the interval $[0,A]$, and therefore, Equation (3.3) is readily applicable. This in turn produces a non-negligible simplification in the corresponding numerics. Finally, we should mention that one of the key characteristics of our approach is its ability to provide integral equations for a multitude of performance measures, including: a) the ARL to false alarm and detection; b) the average detection delay for any arbitrary change-point point $\nu>0$; and c) other performance metrics such as $\RIADD$ and $\STADD$. To the best of our knowledge such pluralism of performance characteristics has never been offered before.

Next we apply the proposed numerical methodology to the Gaussian example and we compare the performance of the two popular tests, namely the CUSUM and Shiryaev-Roberts procedures. We note that it is the first time that such computations are performed for the Shiryaev-Roberts test.

\vskip 3mm

\noindent 4.  AN EXAMPLE

\noindent Consider a Gaussian example of detecting a change in the mean value where
observations are i.i.d. $\mathcal{N}(0,1)$ pre-change and i.i.d. $\mathcal{N}(\theta,1)$, $\theta\neq0$ post-change. Specifically
$$
f(x)=\dfrac{1}{\sqrt{2\pi}}\exp\left\{-\dfrac{x^2}{2}
\right\}\quad\text{and}\quad
g(x)=\dfrac{1}{\sqrt{2\pi}}\exp\left\{-\dfrac{(x-\theta)^2}{2}\right\}.
$$
Recall that we are interested in comparing the operating characteristics of the CUSUM and Shiryaev-Roberts detection procedures expressed via the stationary average detection delay $\STADD(\T)$ on one hand and the supremum average detection delay $\SADD(\T)$ on the other, both as functions of the ARL to false alarm $\EV_\infty[\T]$. As we mentioned before, for both procedures $\SADD(\T)$ coincides with Lorden's essential supremum measure $\mathcal{J}_{\rm L}(\T)$ and with ARL to detection $\EV_0[\T]$. We compute the desired performance metrics for values of the ARL to false alarm $\ARL(\T)=\EV_\infty[\T]$ between 1 and $10^4$ and for characteristic values of the post-change mean $\theta=\{0.01,0.1,0.5,1.0\}$.

Before continuing with the presentation of our numerical results, it is worth mentioning that in order to evaluate the ARL to false alarm of the CUSUM and Shiryaev-Roberts procedures, it is important to obtain preliminary estimates of the threshold $A$ to narrow the domain of search for satisfying the false alarm constraint with equality. For CUSUM we used the following first-order approximation
$$
\ARL(\T_{A}^{\scriptscriptstyle\mathrm{CS}})\approx 2A/(\theta v^2),
$$
which follows from Tartakovsky (2005), where constant $0<v<1$ is the subject of renewal theory. For the Gaussian model considered this constant can be computed numerically as
$$
v=\frac{2}{\theta^2}\exp\left\{-2\sum_{k=1}^\infty\frac{1}{k}\Phi\left(-\frac{\theta}{2}\sqrt{k}\right)\right\},
$$
where
$$
\Phi(x)=\frac{1}{\sqrt{2\pi}}\int_{-\infty}^x e^{-t^2/2} \, dt
$$
is the standard normal distribution function. Also, for small values of $\theta$ Siegmund's corrected Brownian motion approximations are fairly accurate (cf. Siegmund, 1985). For the Shiryaev-Roberts procedure, we used the following approximation due to Pollak (1987):
$$
\ARL(\T_{A}^{\scriptscriptstyle\mathrm{SR}})\approx A/v,
$$
which is very accurate even for relatively small threshold values ($A \ge 20$).

\begin{figure}[p]
    \centering
    \subfigure[$\STADD(\T)$ vs. $\ARL(\T)$]{\label{fig:RIADD-ARL-0_01}
        \includegraphics[width=0.475\textwidth]{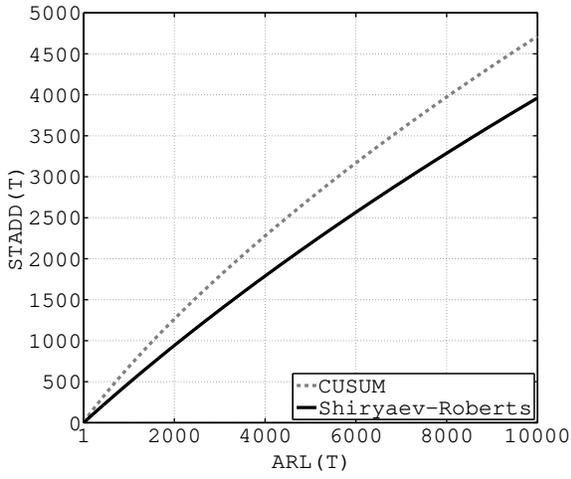}
    }
    \subfigure[$\SADD(\T)$ vs. $\ARL(\T)$]{\label{fig:SADD-ARL-0_01}
        \includegraphics[width=0.475\textwidth]{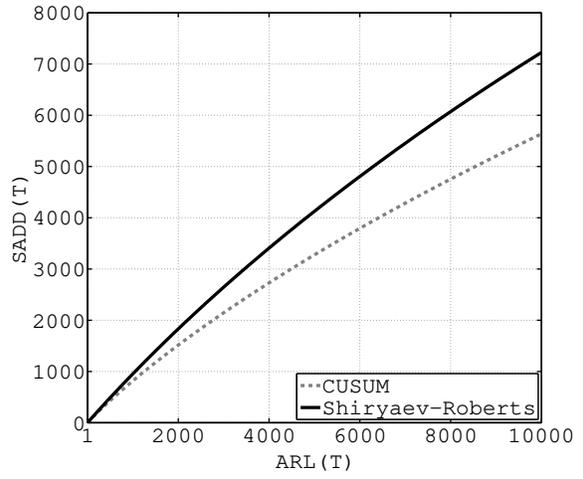}
    }
    \caption{Operating characteristics of CUSUM and Shiryaev-Roberts procedures for $\theta=0.01$.}
    \label{fig:RIADD-SADD-ARL-0_01}
\end{figure}

\begin{figure}[p]
    \centering
    \subfigure[$\STADD(\T)$ vs. $\ARL(\T)$]{\label{fig:RIADD-ARL-0_1}
        \includegraphics[width=0.475\textwidth]{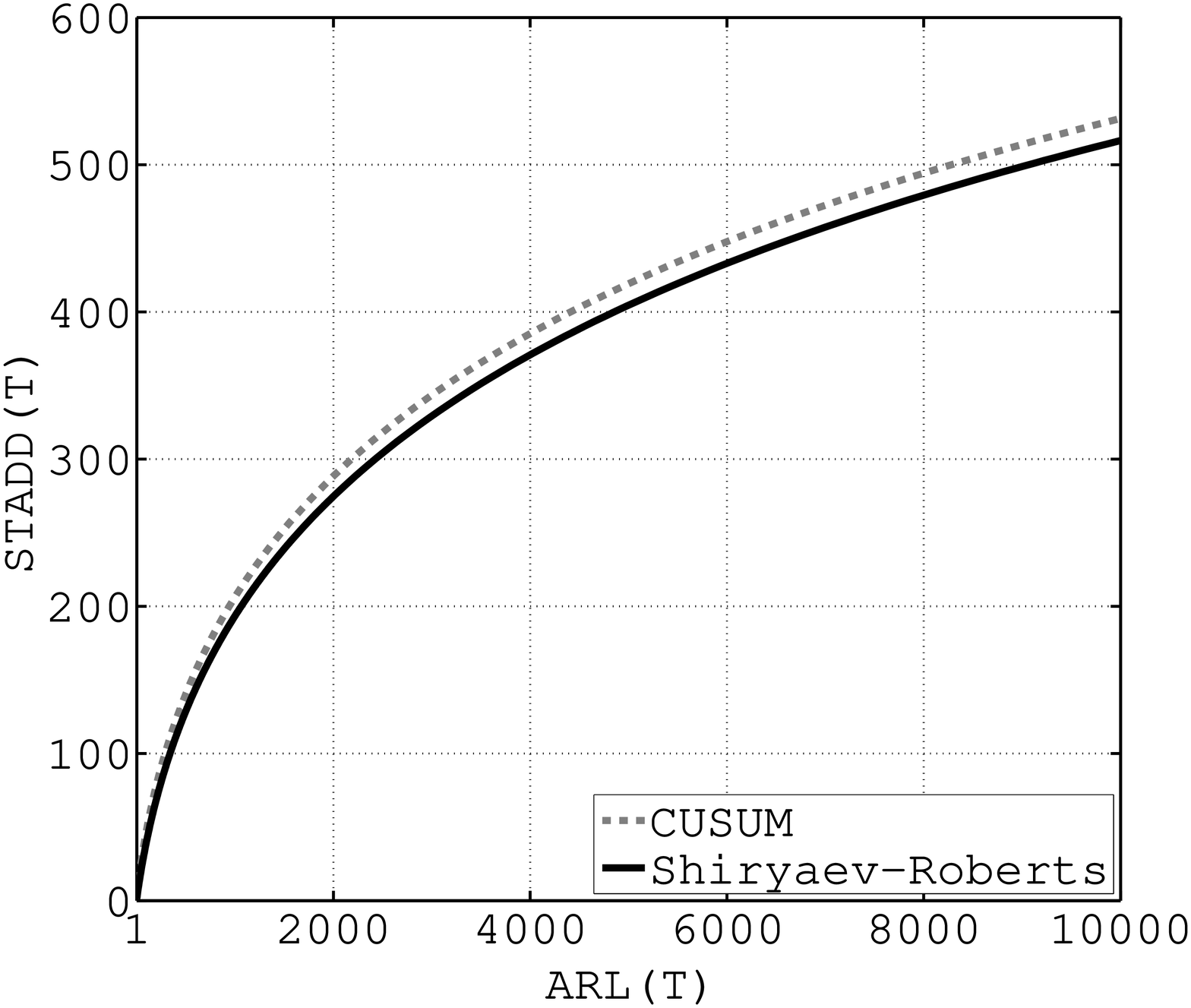}
    }
    \subfigure[$\SADD(\T)$ vs. $\ARL(\T)$]{\label{fig:SADD-ARL-0_1}
        \includegraphics[width=0.475\textwidth]{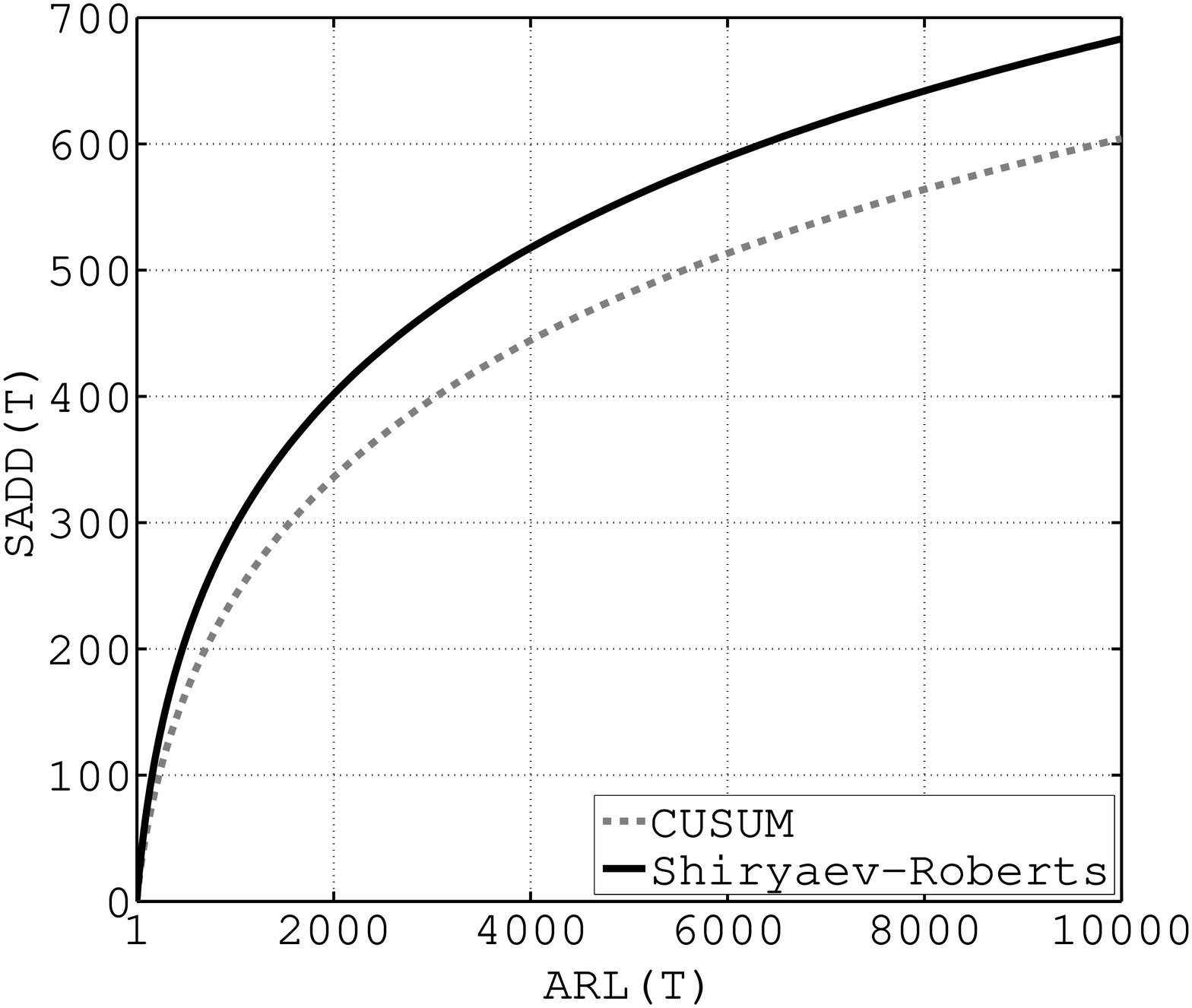}
    }
    \caption{Operating characteristics of CUSUM and Shiryaev-Roberts procedures for $\theta=0.1$.}
    \label{fig:RIADD-SADD-ARL_0_1}
\end{figure}

\begin{figure}[p]
    \centering
    \subfigure[$\STADD(\T)$ vs. $\ARL(\T)$]{\label{fig:RIADD-ARL-0_5}
        \includegraphics[width=0.475\textwidth]{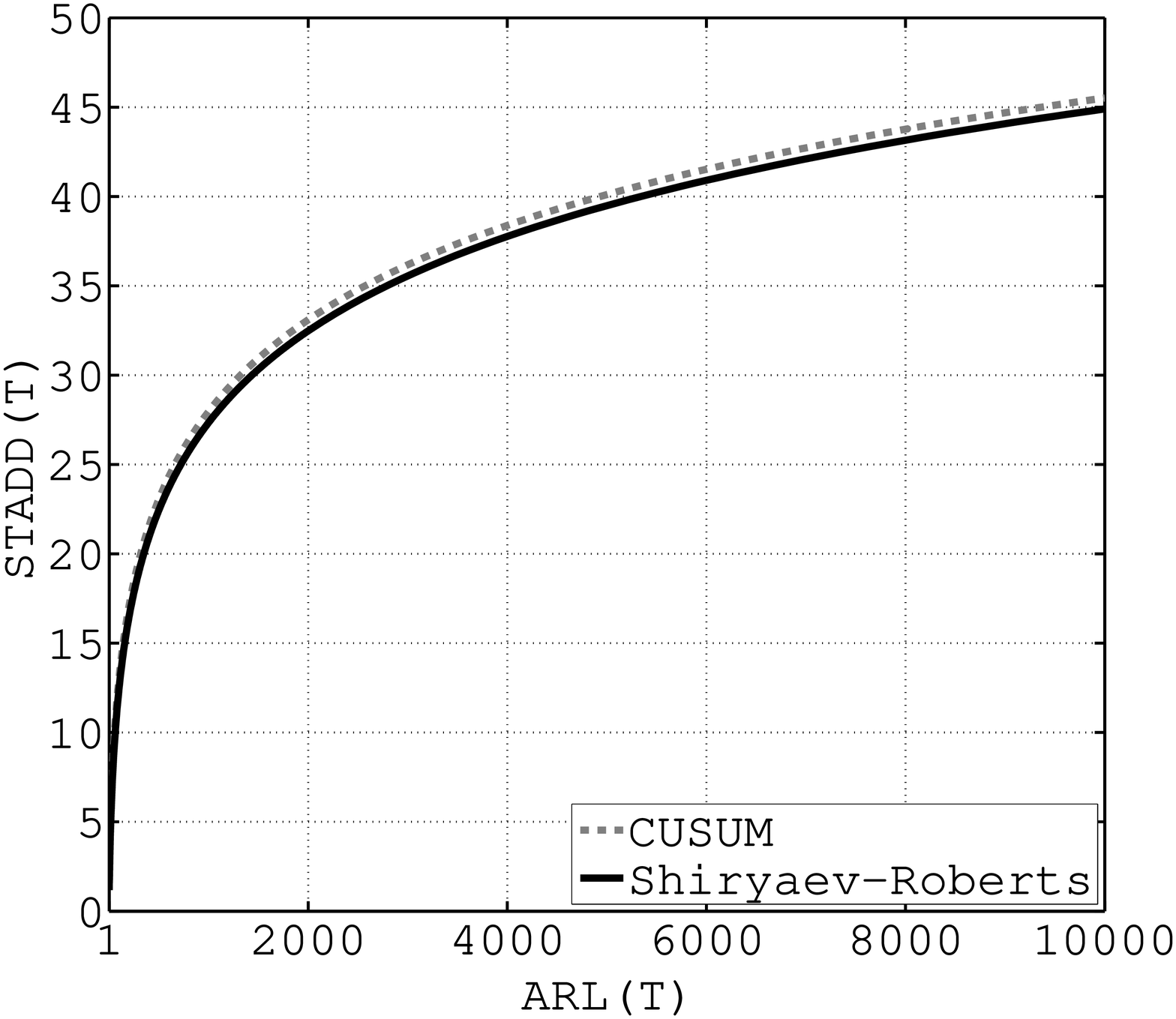}
    }
    \subfigure[$\SADD(\T)$ vs. $\ARL(\T)$]{\label{fig:SADD-ARL-0_5}
        \includegraphics[width=0.475\textwidth]{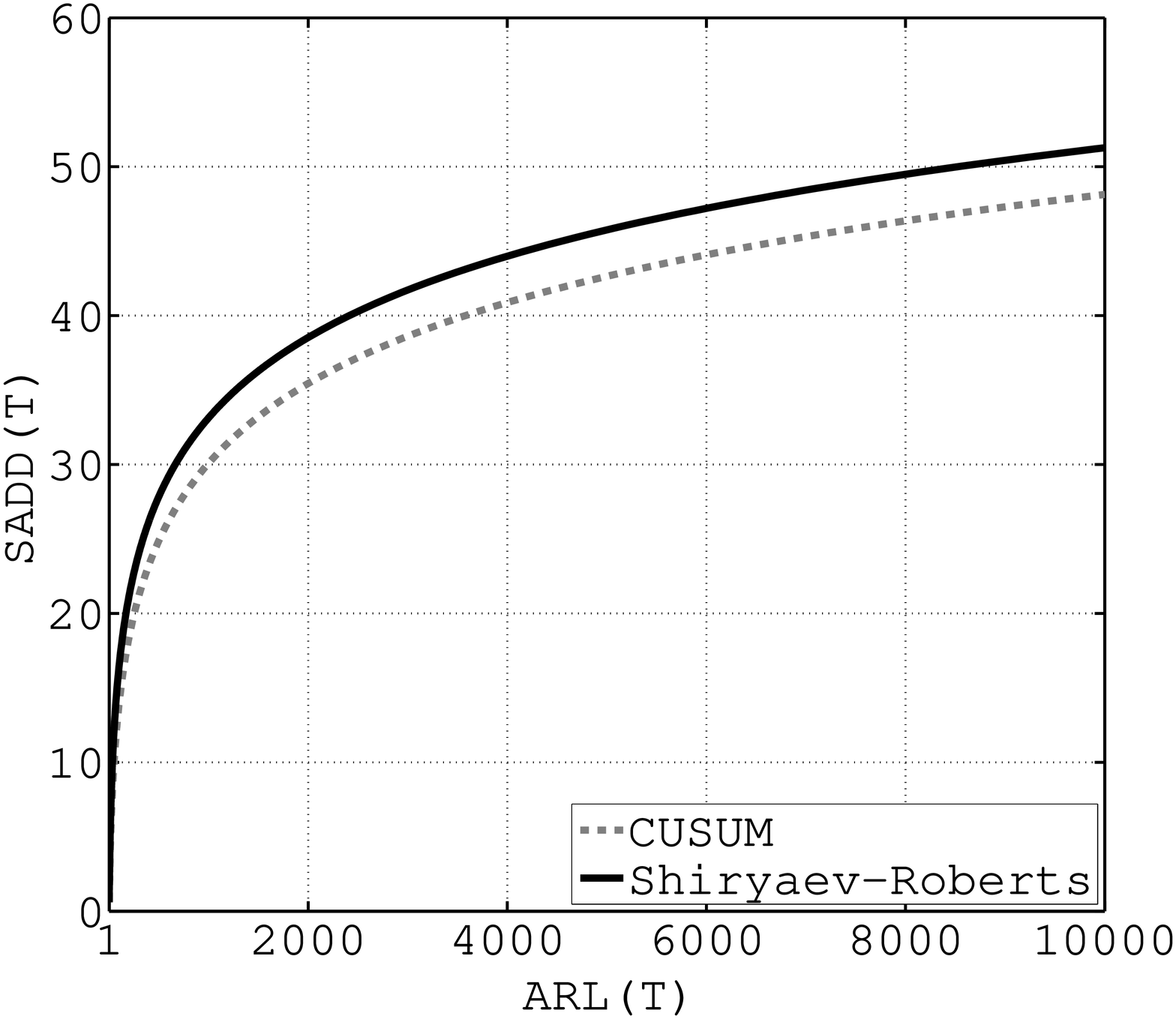}
    }
    \caption{Operating characteristics of CUSUM and Shiryaev-Roberts procedures for $\theta=0.5$.}
    \label{fig:RIADD-SADD-ARL_0_5}
\end{figure}

\begin{figure}[p]
    \centering
    \subfigure[$\STADD(\T)$ vs. $\ARL(\T)$]{\label{fig:RIADD-ARL-1_0}
        \includegraphics[width=0.475\textwidth]{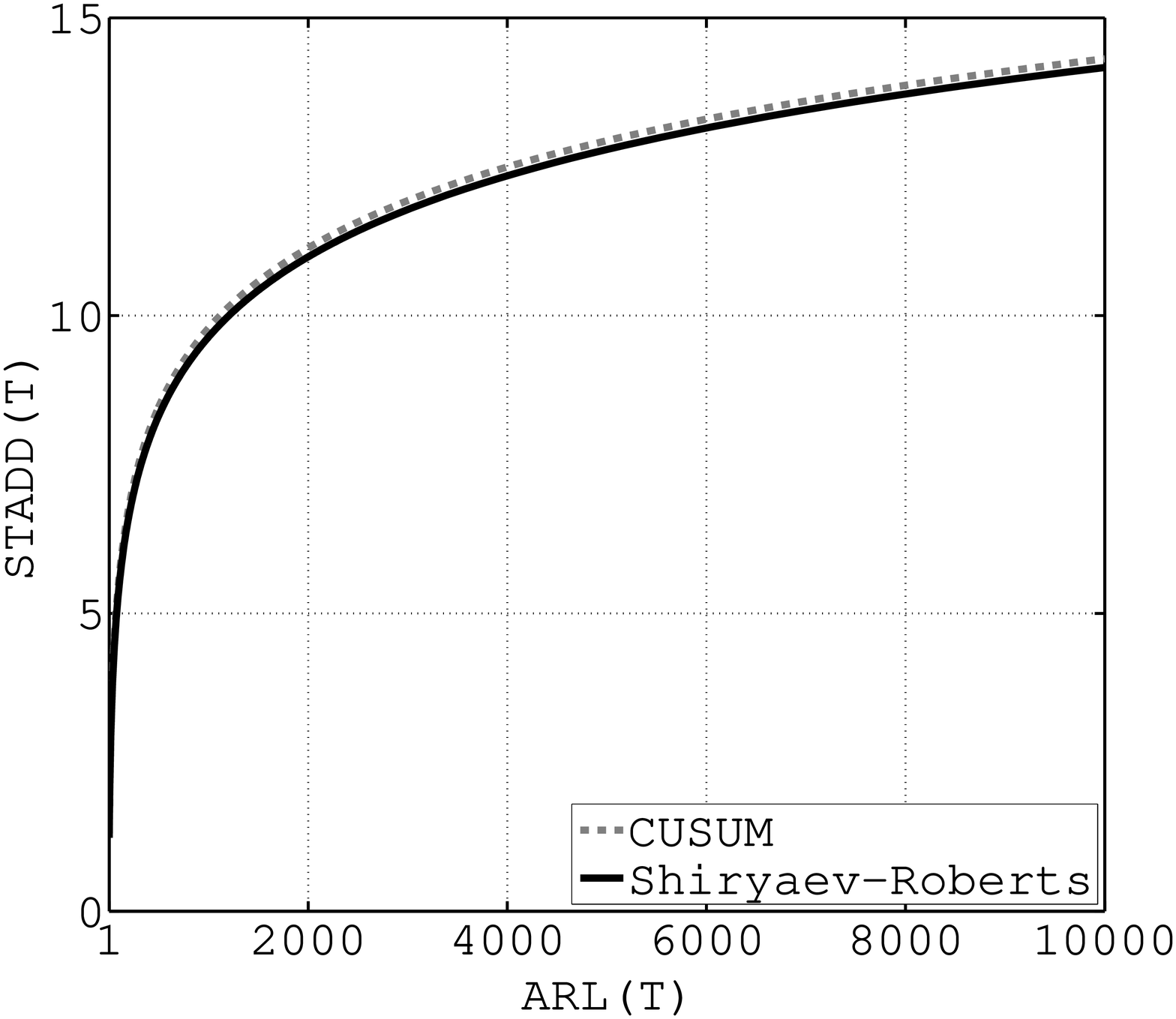}
    }
    \subfigure[$\SADD(\T)$ vs. $\ARL(\T)$]{\label{fig:SADD-ARL-1_0}
        \includegraphics[width=0.475\textwidth]{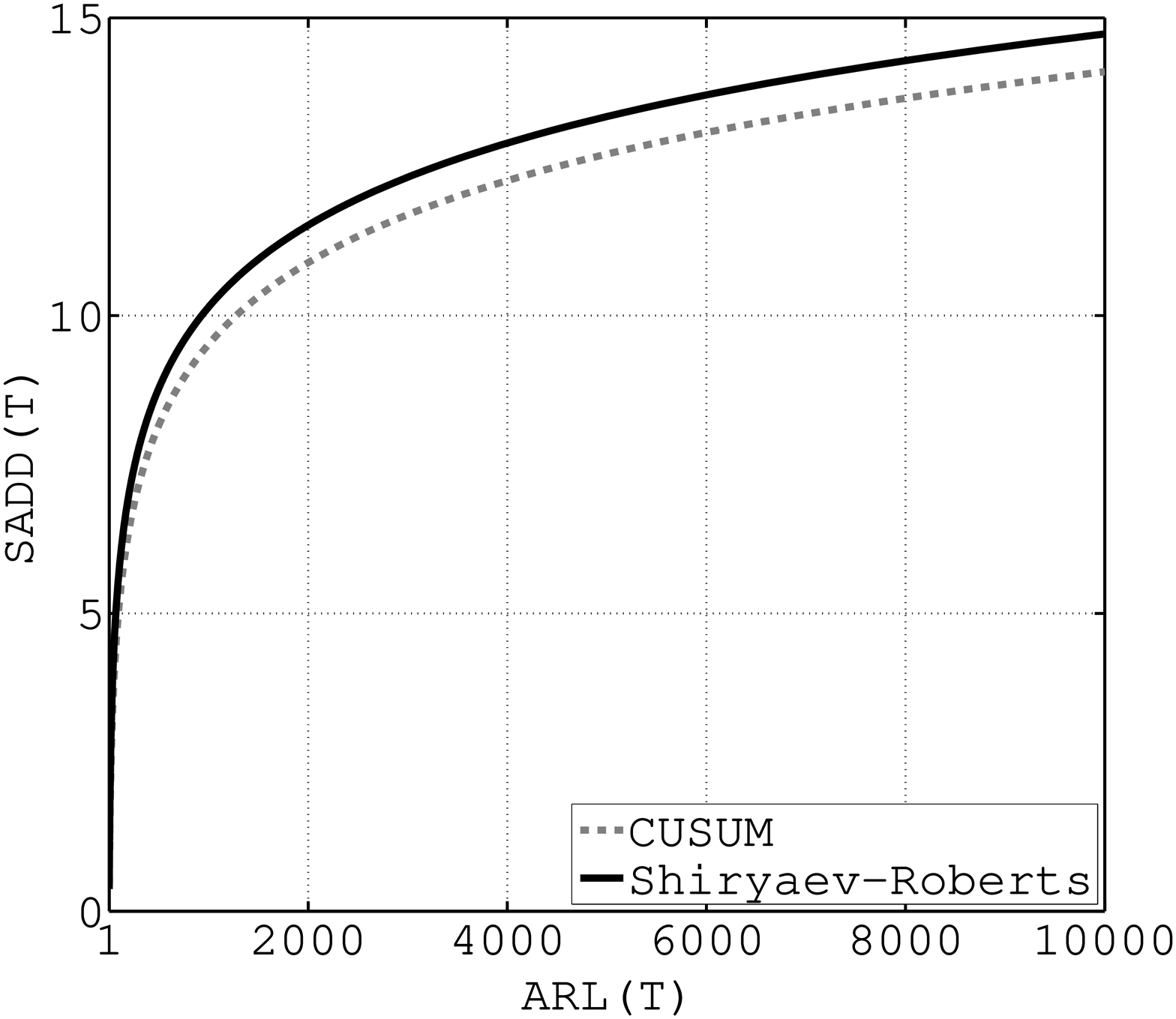}
    }
    \caption{Operating characteristics of CUSUM and Shiryaev-Roberts procedures for $\theta=1.0$.}
    \label{fig:RIADD-SADD-ARL_1_0}
\end{figure}

Figures \ref{fig:RIADD-SADD-ARL-0_01}--\ref{fig:RIADD-SADD-ARL_1_0} and Tables \ref{tab:summary_0_01}--\ref{tab:summary_1_0} show the operating characteristics for the aforementioned set of parameters.  As expected, the CUSUM procedure outperforms the Shiryaev-Roberts procedure in the minimax scenario. The Shiryaev-Roberts procedure, on the other hand, performs better with respect to the stationary average detection delay for detecting distant changes using a repeated application of the same stopping rule. As we can see, the difference is significant only for small changes, visible for moderate changes, while the two  procedures perform equally well for large changes.

The precision of our numerical approximations was verified by using Monte Carlo techniques for several parameter values. In all cases, the difference was negligible. We also note that for the Gaussian example considered in this section, Dragalin (1994) proposed a different, more accurate but also computationally more demanding method for computing the ARL to false alarm $\EV_\infty[\T_{A}^{\scriptscriptstyle\mathrm{CS}}]$ and the ARL to detection $\EV_0[\T_{A}^{\scriptscriptstyle\mathrm{CS}}]$ of the CUSUM procedure. Comparing our results with the outcome of this more complex approach shows that the difference is very small. This fact is an additional indication that our simple numerical method is of sufficiently high accuracy.
\vskip 3mm

\noindent ACKNOWLEDGEMENTS

\noindent This work was supported in part by the U.S.\ Army Research Office MURI grant
W911NF-06-1-0094 and by the U.S.\ National Science Foundation grant CCF-0830419 at
the University of Southern California. We are grateful to the reviewer for valuable suggestions.
\vskip 3mm

\noindent BIBLIOGRAPHY

\noindent Atkinson, K., and Han, W. (2001). {\it Theoretical Numerical Analysis: A Functional Analysis Framework}. New York: Springer-Verlag.
\vskip 3mm

\noindent Dragalin, V. V. (1994). Optimal CUSUM envelope for monitoring the mean of normal distribution. {\it Economic Quality Control}, {\bf 9}, 185--202.
\vskip 3mm

\noindent Feinberg, E. A., and Shiryaev, A. N. (2006). Quickest detection of drift change for Brownian motion in generalized Bayesian and minimax settings. {\it Statistics \& Decisions}, {\bf 24}, 445--470.
\vskip 3mm

\noindent Kantorovich, L. V., and Krylov, V. I. (1958). {\it Approximate Methods of Higher Analysis}. New York: Interscience Publishers, Inc.
\vskip 3mm

\noindent Kress, R. (1989). {\it Linear Integral Equations}. Berlin: Springer-Verlag.
\vskip 3mm

\noindent Lai, T. L. (1998). Information bounds and quick detection of parameter changes in stochastic systems. {\it IEEE Transactions on Information Theory}, {\bf 44}, 2917--2929.
\vskip 3mm

\noindent Lorden, G. (1971). Procedures for reacting to a change in distribution. {\it Annals of Mathematical Statistics}, {\bf 42}, 1897--1908.
\vskip 3mm

\noindent Moustakides, G. V. (1986). Optimal stopping times for detecting changes in distributions. {\it Annals of Statistics}, {\bf 14}, 1379--1387.
\vskip 3mm

\noindent Moustakides, G. V. (2008). Sequential change detection revisited. {\it Annals of Statistics}, {\bf 36}, 1452--1465.
\vskip 3mm

\noindent Moustakides, G. V., Polunchenko, A. S., and Tartakovsky, A. G. (2009). A numerical approach to comparative efficiency analysis of quickest change-point detection procedures. {\it Statistica Sinica}, submitted.
\vskip 3mm

\noindent Page, E. S. (1954). Continuous inspection schemes. {\it Biometrika}, {\bf 41}, 100--115.
\vskip 3mm

\noindent Petrovskii, I. G. (1957). {\it Lectures on the Theory of Integral Equations}. New-York: Graylock Press, Rochester.
\vskip 3mm

\noindent Pollak, M. (1985). Optimal detection of a change in distribution. {\it Annals of Statistics}, {\bf 13}, 206--227.
\vskip 3mm

\noindent Pollak, M. (1987). Average run lengths of an optimal method of detecting a change in distribution. {\it Annals of Statistics}, {\bf 15}, 749--779.
\vskip 3mm

\noindent Pollak, M., and Siegmund, D. (1985). A diffusion process and its applications to detecting a change in the drift of Brownian motion. {\it Biometrika}, {\bf 72}, 267--280.
\vskip 3mm

\noindent Pollak, M., and Tartakovsky, A. G. (2009). Optimality properties of the Shiryaev-Roberts procedure. {\it Statistica Sinica}, in press.
\vskip 3mm

\noindent Ritov, Y. (1990). Decision theoretic optimality of the CUSUM procedure. {\it Annals of Statistics}, {\bf 18}, 1466--1469.
\vskip 3mm

\noindent Roberts, S. W. (1966). A comparison of some control chart procedures. {\it Technometrics}, {\bf 8}, 411--430.
\vskip 3mm

\noindent Shiryaev, A. N. (1961). The problem of the most rapid detection of a disturbance in a stationary process. {\it Soviet Math.\ Dokl.}, {\bf 2}, 795--799 (translation from Dokl. Akad. Nauk SSSR {\bf 138}, 1039--1042, 1961).
\vskip 3mm

\noindent Shiryaev, A. N. (1963). On optimum methods in quickest detection problems. {\it Theory Probability and Its Applications}, {\bf 8}, 22--46.
\vskip 3mm

\noindent Siegmund, D. (1985). {\it Sequential Analysis: Tests and Confidence Intervals}. New York: Springer-Verlag.
\vskip 3mm

\noindent Srivastava, M. S., and Wu, Y. (1993). Comparison of EWMA, CUSUM and Shiryayev-Roberts procedures for detecting a shift in the mean. {\it Annals of Statistics}, {\bf 21}, 645--670.
\vskip 3mm

\noindent Tartakovsky, A. G. (2005). Asymptotic performance of a multichart CUSUM test under false alarm probability constraint. {\it Proceedings of 44th IEEE Conference on Decision and Control and the European Control Conference (CDC-ECC'05)}, Seville, Spain, Omnipress CD-ROM, ISBN 0-7803-9568-9, 320--325.
\vskip 3mm

\noindent Tartakovsky, A. G., and Ivanova, I. A. (1992). Comparison of some sequential rules for detecting changes in distributions. {\it Problems of Information Transmission}, {\bf 28}, 117--124.
\vskip 3mm

\noindent Tartakovsky, A. G., and Veeravalli, V. V. (2004). Change-point detection in multichannel and distributed systems with applications. Applications of
Sequential Methodologies, N.\ Mukhopadhyay, S. Datta and S. Chattopadhyay, eds., pp. 339--370, New York: Marcel Dekker.
\vskip 3mm

\begin{table}[p]
    \centering
    \caption{Operating characteristics of CUSUM and Shiryaev-Roberts procedures for $\theta=0.01$}
        \begin{tabular}{|l||c||c|c|c|c|c|c|}
    \hline Test &$\gamma$ &50 &100 &500 &1000 &5000 &10000\\
        \hline
    \hline
        \multirow{4}[8]{3.15cm}{CUSUM}
        &$A$ &1.06 &1.091 &1.2263 &1.3348 &1.861 &2.3304 \bigstrut\\\cline{2-8}
        &$\ARL$ &50.05 &100.8 &500.37 &1000.2 &5000.8 &10000.12 \bigstrut\\\cline{2-8}
        &$\STADD$ &40.31 &79.14 &361.68 &682.9 &2736.65 &4712.65 \bigstrut\\\cline{2-8}
        &$\SADD$ &47.77 &94.38 &433.36 &818.6 &3277.69 &5636.54\bigstrut\\
        \hline
    \hline
        \multirow{4}[8]{3.15cm}{Shiryaev-Roberts}
        &$A$ &49.71 &99.42 &497.1 &994.19 &4970.95 &9941.91 \bigstrut\\\cline{2-8}
        &$\ARL$ &50.33 &100.29 &500.26 &1000.25 &5000.2 &10000.15 \bigstrut\\\cline{2-8}
        &$\STADD$ &25.62 &50.48 &246.6 &485.06 &2186.23 &3961.42 \bigstrut\\\cline{2-8}
        &$\SADD$ &50.21 &99.79 &488.32 &954.57 &4126.98 &7226.55 \bigstrut\\
    \hline
    \end{tabular}
    \label{tab:summary_0_01}
\end{table}

\begin{table}[p]
    \centering
    \caption{Operating characteristics of CUSUM and Shiryaev-Roberts procedures for $\theta=0.1$}
    \begin{tabular}{|l||c||c|c|c|c|c|c|}
    \hline Test &$\gamma$ &50 &100 &500 &1000 &5000 &10000\\
        \hline
    \hline
        \multirow{4}[8]{3.15cm}{CUSUM}
        &$A$ &1.676 &2.1 &4.575 &7.205 &26.15 &48.964\bigstrut\\\cline{2-8}
        &$\ARL$ &50.03 &100.2 &500.64 &1000.8 &5000.1 &10000.62\bigstrut\\\cline{2-8}
        &$\STADD$ &27.81 &47.6 &140.52 &206.4 &419.2 &531.48\bigstrut\\\cline{2-8}
        &$\SADD$ &32.8 &56.45 &166.34 &242.97 &482.88 &605.15\bigstrut\\
        \hline
    \hline
        \multirow{4}[8]{3.15cm}{Shiryaev-Roberts}
        &$A$ &47.17 &94.34 &471.7 &943.41 &4717.04 &9434.08 \bigstrut\\\cline{2-8}
        &$\ARL$ &50.29 &100.28 &500.28 &1000.28 &5000.24 &10000.17 \bigstrut\\\cline{2-8}
        &$\STADD$ &22.43 &40.14 &128.85 &193.5 &404.58 &516.46 \bigstrut\\\cline{2-8}
        &$\SADD$ &41.4 &72.32 &209.44 &298.5 &557.87 &684.17\bigstrut\\
    \hline
    \end{tabular}
    \label{tab:summary_0_1}
\end{table}

\begin{table}[p]
    \centering
    \caption{Operating characteristics of CUSUM and Shiryaev-Roberts procedures for $\theta=0.5$}
    \begin{tabular}{|l||c||c|c|c|c|c|c|}
    \hline Test &$\gamma$ &50 &100 &500 &1000 &5000 &10000\\
        \hline
    \hline
        \multirow{4}[8]{3.15cm}{CUSUM}
        &$A$ &5.45 &9.15 &37.88 &73.2 &353.58 &703.78 \bigstrut\\\cline{2-8}
        &$\ARL$ &50.76 &99.57 &499.42 &999.69 &4999.38 &9999.21 \bigstrut\\\cline{2-8}
        &$\STADD$ &9.69 &13.03 &23.05 &27.96 &40.1 &45.51 \bigstrut\\\cline{2-8}
        &$\SADD$ &10.56 &14.37 &25.37 &30.58 &43.13 &48.63 \bigstrut\\
        \hline
    \hline
        \multirow{4}[8]{3.15cm}{Shiryaev-Roberts}
        &$A$ &37.38 &74.76 &373.81 &747.62 &3738.08 &7476.15 \bigstrut\\\cline{2-8}
        &$\ARL$ &50.44 &100.44 &500.45 &1000.45 &5000.45 &10000.24 \bigstrut\\\cline{2-8}
        &$\STADD$ &9.08 &12.49 &22.45 &27.35 &39.49 &44.9 \bigstrut\\\cline{2-8}
        &$\SADD$ &13.09 &17.39 &28.84 &34.13 &46.76 &52.27 \bigstrut\\
    \hline
    \end{tabular}
    \label{tab:summary_0_5}
\end{table}

\begin{table}[p]
    \centering
    \caption{Operating characteristics of CUSUM and Shiryaev-Roberts procedures for  $\theta=1.0$}
    \begin{tabular}{|l||c||c|c|c|c|c|c|}
    \hline Test &$\gamma$ &50 &100 &500 &1000 &5000 &10000\\
        \hline
    \hline
        \multirow{4}[8]{3.15cm}{CUSUM}
        &$A$ &9.32 &17.33 &80.65 &159.35 &788.0 &1574.0 \bigstrut\\\cline{2-8}
        &$\ARL$ &49.43 &99.33 &499.5 &999.39 &4999.25 &9999.38\bigstrut\\\cline{2-8}
        &$\STADD$ &4.48 &5.59 &8.47 &9.79 &12.94 &14.31 \bigstrut\\\cline{2-8}
        &$\SADD$ &4.63 &5.85 &8.89 &10.25 &13.45 &14.83 \bigstrut\\
        \hline
    \hline
        \multirow{4}[8]{3.15cm}{Shiryaev-Roberts}
        &$A$ &28.02 &56.04 &280.19 &560.37 &2801.75 &5603.7 \bigstrut\\\cline{2-8}
        &$\ARL$ &50.79 &100.79 &500.8 &1000.79 &5001.75 &10000.86 \bigstrut\\\cline{2-8}
        &$\STADD$ &4.37 &5.46 &8.33 &9.64 &12.79 &14.17 \bigstrut\\\cline{2-8}
        &$\SADD$ &5.46 &6.71 &9.78 &11.14 &14.34 &15.73 \bigstrut\\
    \hline
    \end{tabular}
    \label{tab:summary_1_0}
\end{table}

\end{document}